# MODELS OF HIGH-LUMINOSITY ACCRETION DISKS SURROUNDING BLACK HOLES

A. S. Klepnev[1,2] and G. S. Bisnovatyi-Kogan[1,2]

*The problem of steady-state accretion to nonrotating black holes is examined. Advection is included and generalized formulas for the radiation pressure in both the optically thick and thin cases are used. Special attention is devoted to models with a high accretion rate. Global solutions for accretion disks are studied which describe a continuous transition between an optically thick outer region and an optically thin inner region. It is shown that there is a maximum disk temperature for the model with a viscosity parameter α = 0.5. For the model with α = 0.1, no optically thin regions are found to exist for any accretion rate.*

## 1. Introduction

The standard model for accretion disks [1] is based on several serious simplifying assumptions. The disk must be geometrically thin and rotate at the Kepler angular velocity. These assumptions make it possible to neglect radial gradients and, ultimately, to proceed from the differential equations to algebraic equations. For low accretion rates $\dot{M}$, this assumption is fully appropriate.

However, since the end of the 1970's it has been demonstrated that for high accretion rates, the disk structure may differ from the standard model. To solve the more general problem, advection and a radial pressure gradient have been included in the analysis of the disk structure [2]. Numerical solutions for advective accretion disk have been obtained in the optically thick case [3,4].

———

Local solutions of the equations without advection and including general formulas for the radiative emission and radiative pressure reveal the existence of two types of solutions: optically thick and optically thin, which do not overlap if $\dot{M} < \dot{M}_{cr} \approx (0.6 - 0.9)\dot{M}_{EDD}$ for $\alpha = 1$ and $M_{BH} = 10^8 M_\odot$ [5]. It was shown that for large accretion rates there are no local solutions that are continuous over the entire region of existence of the disk and undergo Kepler rotation. This was explained [5] by the fact that advection begins to play an important role for accretion rates greater than $\dot{M}_{cr}$.

A self-consistent solution for an advective accretion disk with a continuous description of the entire region between the optically thin and optically thick regions has been obtained [6]. Here we construct a solution for an advective accretion disk under the same physical assumptions as in Ref. 6 but with wider ranges of the parameters $\alpha$ and $\dot{M}$. A more accurate expression [7] is used for the effective optical thickness, $\tau_* = ((\tau_0 + \tau_\alpha)\tau_\alpha)^{1/2}$. A new program for the numerical calculations based on a relaxation method has been written, while the inner singularity is studied using the method described in Appendix B of Ref. 4.

## 2. Basic equations

We consider the equations for a one-dimensional, steady-state accretion disk which is averaged in the vertical direction. These equations include advection and can be used for any value of the vertical optical thickness of the disk. We use a pseudo-newtonian approximation for the structure of the disk near the black hole, where the effects of the general theory of relativity are taken into account using the Paczyñski-Wiita potential [8]

$$\Phi(r) = -\frac{GM}{r - 2r_g}. \tag{1}$$

Here $M$ is the mass of the black hole and $2r_g = 2GM/c^2$ is the gravitational radius. The self gravitation of the disk is neglected.

In our model we use the simplest parametrization of the viscosity tensor [1],

$$t_{r\varphi} = -\alpha P. \tag{2}$$

The conservation of mass is expressed in the form

$$\dot{M} = 4\pi r h \rho v, \tag{3}$$

where $\dot{M}$ is the accretion rate, $\dot{M} > 0$, and $h$ is the half thickness of the disk.

The equilibrium in the vertical direction is described by

$$\frac{dP}{dz} = -\rho z \Omega_K^2. \tag{4}$$

Replacing the derivative on the left by the ratio $P/h$ and assuming $z = h$ on the right, this equation can be written for the case of a geometrically thin disk as

$$h = \frac{c_s}{\Omega_K}, \tag{5}$$

where $c_s = \sqrt{P/\rho}$ is the isothermal sound speed.

The equations of motion in the radial and azimuthal directions are

$$v\frac{dv}{dr} = -\frac{1}{\rho}\frac{dP}{dr} + \left(\Omega^2 - \Omega_K^2\right)r, \tag{6}$$

and

$$\frac{\dot{M}}{4\pi}\frac{d\ell}{dr} + \frac{d}{dr}\left(r^2 h t_{r\varphi}\right) = 0 \tag{7}$$

respectively, where $\Omega_K$ is the Kepler angular velocity, given by $\Omega_K^2 = GM/r(r-2r_g)^2$, $\ell = \Omega r^2$ is the specific angular momentum, and $t_{r\varphi}$ is the $(r,\varphi)$ component of the viscosity tensor. The other components of this tensor are assumed negligibly small. The vertically averaged equation for the conservation of energy is

$$Q_{adv} = Q^+ - Q^-, \tag{8}$$

where

$$Q_{adv} = -\frac{\dot{M}}{4\pi r}\left[\frac{dE}{dr} + P\frac{d}{dr}\left(\frac{1}{\rho}\right)\right], \tag{9}$$

$$Q^+ = -\frac{\dot{M}}{4\pi}r\Omega\frac{d\Omega}{dr}\left(1 - \frac{l_{in}}{l}\right), \tag{10}$$

$$Q^- = \frac{2aT^4 c}{3(\tau_\alpha + \tau_0)h}\left[1 + \frac{4}{3(\tau_0 + \tau_\alpha)} + \frac{2}{3\tau_*^2}\right]^{-1} \tag{11}$$

are the energy fluxes (erg/cm²/s) associated with advection, viscous dissipation, and radiation from the surface, respectively. Here $T$ is the temperature, $a$ is the constant of the radiation energy density, and $\tau_0$ is the Thomson optical depth, given by $\tau_0 = 0.4\rho h$ for a hydrogenic composition. Here we have introduced the optical thickness for absorption,

$$\tau_\alpha = 5.2 \cdot 10^{21} \frac{\rho^2 T^{1/2} h}{acT^4}, \tag{12}$$

and the effective optical thickness

$$\tau_* = \left[(\tau_0 + \tau_\alpha)\tau_\alpha\right]^{1/2}. \tag{13}$$

Here we use the formula for the effective optical thickness $\tau_*$ in its full form, as opposed to the earlier work [4], where the approximation $\tau_* = (\tau_0 \tau_\alpha)^{1/2}$ was used in light of the inequality $\tau_0 \gg \tau_\alpha$.

The equation of state is for a mixture of matter and radiation, with

$$P_{tot} = P_{gas} + P_{rad}. \tag{14}$$

Here the gas pressure is given by the standard formula,

$$P_{gas} = \rho RT, \tag{15}$$

where $R$ is the gas constant.

The radiation pressure is given by

$$P_{rad} = \frac{aT^4}{3}\left[1 + \frac{4}{3(\tau_0 + \tau_\alpha)}\right]\left[1 + \frac{4}{3(\tau_0 + \tau_\alpha)} + \frac{2}{3\tau_*^2}\right]^{-1}. \tag{16}$$

The specific energy of the mixture of matter and radiation is

$$\rho E = \frac{3}{2} P_{gas} + 3 P_{rad}. \tag{17}$$

Expressions for $Q^-$ and $P_{rad}$ which are valid for any optical thickness have been obtained for the case $\tau_0 \gg \tau_\alpha$ in Ref. 5.

## 3. Singular points

According to Ref. 6, this system of differential and algebraic equations can be reduced to two ordinary differential equations,

$$\frac{x}{v}\frac{dv}{dx} = \frac{N}{D}, \tag{18}$$

$$\frac{x}{v}\frac{dc_s}{dx} = 1 - \left(\frac{v^2}{c_s^2} - 1\right)\frac{N}{D} + \frac{x^2}{c_s^2}\left(\Omega^2 - \frac{1}{x(x-2)^2}\right) + \frac{3x-2}{2(x-2)}. \tag{19}$$

Here the numerator $N$ and denominator $D$ are algebraic expressions depending on $x$, $v$, $c_s$, and $l_{in}$ and these equations have been written in dimensionless form. The coordinate $x = r/r_g$, where $r_g = GM/c^2$. The velocities v and $c_s$ have been written in dimensionless form in terms of the speed of light, and the specific angular momentum $l_{in}$, in terms of $c/r_g$.

This system of differential equations has two singular points, defined by the conditions

$$D = 0, \quad N = 0. \tag{20}$$

The inner singularity lies near the last stable orbit with $r = 6 r_g$. The outer singularity, lying at distances much greater than $r_g$, is an artifact arising from our use of the artificial parametrization $t_{r\varphi} = -\alpha P$ of the viscosity tensor. If the physically correct parametrization, $t_{r\varphi} = \rho v r \frac{d\Omega}{dr}$, is used then there is no outer singularity [4].

**4. Method of solution**

The system of ordinary differential equations was solved by a finite difference method discussed in Refs. 9 and 4. The method is based on reducing the system of differential equations to a system of nonlinear algebraic equations which are solved by an iterative Newton-Raphson scheme. The most important components of this method are the addition of an expansion of the solution near the inner singularity and the use of $l_{in}$ as an independent variable in the iterative scheme [4].

Our solution is independent of the outer boundary condition. Formally, the numerical method we have used requires a specification of outer boundary conditions for the variables $v$ and $c_s$. However, it was found [4] that this method converges rapidly to a separatrix that passes through the singularities. Only near the outer boundary (at several grid points) will the solution depend on the outer conditions.

**5. Numerical solutions**

We have obtained numerical solutions for the structure of an accretion disk over a wide range of the parameters $\dot{m}$ ($\dot{m} = \dot{M}_c^2/L_{EDD}$) and $\alpha$.

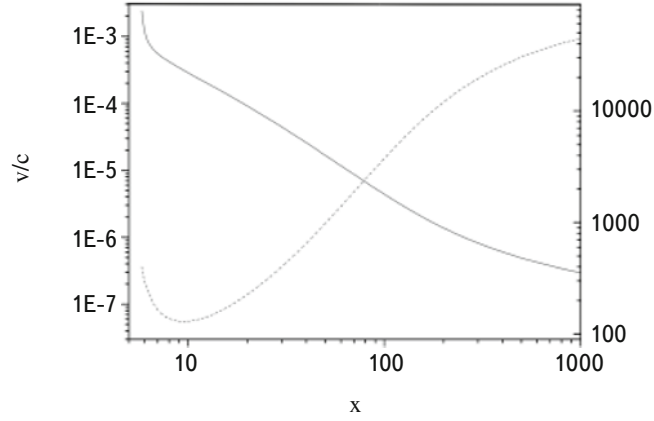

Fig. 1. The radial velocity $v$ of mass flow (smooth curve) and effective optical depth $\tau_*$ (dashed curve) as functions of radius for an accretion rate $\dot{m} = 10$ and viscosity parameter $\alpha = 0.01$

For low accretion rates, $\dot{m} < 0.1$, the solution for the advection model has $\tau_* \gg 1$, $v \ll c_s$, and an angular velocity close to the Kepler velocity everywhere, except a very thin layer near the inner boundary of the disk. Figure 1 shows the radial variation of the radial velocity and of the effective optical thickness for the model with $\alpha = 0.01$ and $\dot{m} = 10$.

As the accretion rate increases, the situation changes significantly. The changes show up primarily in the inner region of the disk. It is there that advection processes primarily begin to play a significant role.

At supercritical accretion rates the following pattern is observed. Figures 2 and 3 show the radial dependences of the effective optical thickness and temperature of the accretion disk for an accretion rate $\dot{m} = 50$ and different values of the viscosity parameter $\alpha$ = 0.01, 0.1 and 0.4. Clearly, for large $\dot{m}$ and $\alpha$ the inner part of the disk becomes optically thin. Because of this, a sharp increase in the temperature of the accretion disk is observed in this region. For the different values of $\alpha$, a solution was obtained which matched the optically thick outer and optically thin inner regions of disk continuously, similarly to Ref. 6.

Two distinct regions can be seen in the plot of the radial dependence of the temperature of the accretion disk. This is especially noticeable for a viscosity parameter $\alpha = 0.4$, where one can see an inner optically thin region with dominant nonequilibrium radiation pressure $P_{rad}$ and an outer region which is optically thick with dominant equilibrium radiation pressure.

Things are different when the viscosity parameter is small. Figures 4 and 5 show the radial dependences of the effective optical depth and temperature of the accretion disk for various accretion rates and a viscosity $\alpha = 0.1$. These graphs show that only a small (considerably smaller than for $\alpha = 0.4$) inner region becomes optically thin for accretion rates of $\dot{m} \approx 30 - 70$. On the other hand, in the case of $\alpha = 0.01$, there will be no optically thin regions at all.

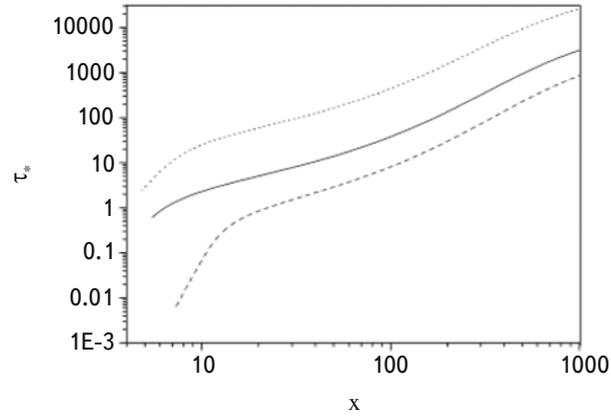

Fig. 2. The radial dependence of the effective optical depth of the accretion disk for an accretion rate $\dot{m}=50$ and viscosity parameters $\alpha=0.01$ (dotted curve), $\alpha=0.1$ (smooth curve), and $\alpha=0.4$ (dashed curve).

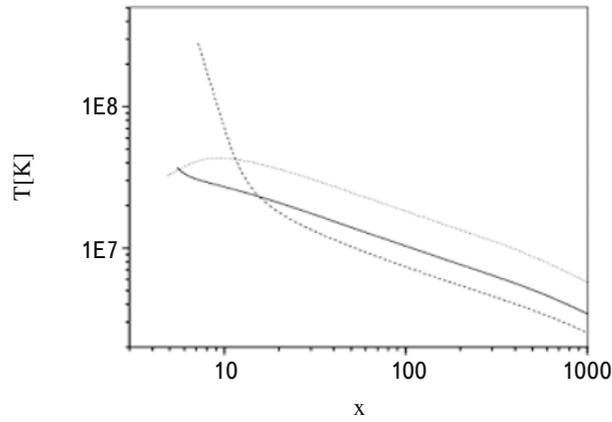

Fig. 3. The radial dependence of the temperature of the accretion disk for an accretion rate $\dot{m}=50$ and viscosity parameters $\alpha=0.01$ (dotted curve), $\alpha=0.1$ (smooth curve), and $\alpha=0.4$ (dashed curve).

Figure 6 shows the radial variation in the half-thickness $h$ of the accretion disk. It shows that for small accretion rates the approximation of a geometrically thin disk ($h<<r$) is well satisfied for all regions of the disk. For high $\dot{m}$, this condition is not as well satisfied, but it is still possible to use this approximation. As this figure shows, the thickness of the disk depends on the accretion rate and, more weakly, on the viscosity parameter $\alpha$.

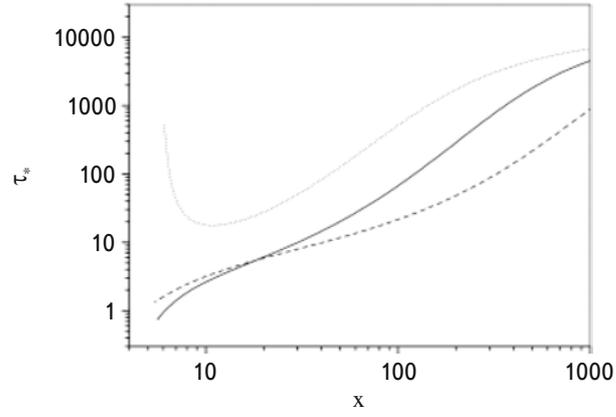

Fig. 4. The radial dependence of the effective optical depth $\tau_e$ of the accretion disk for accretion rates $\dot{m} = 8$ (dotted curve), $\dot{m} = 30$ (smooth curve), and $\dot{m} = 150$ (dashed curve), and a viscosity parameter $\alpha = 0.1$.

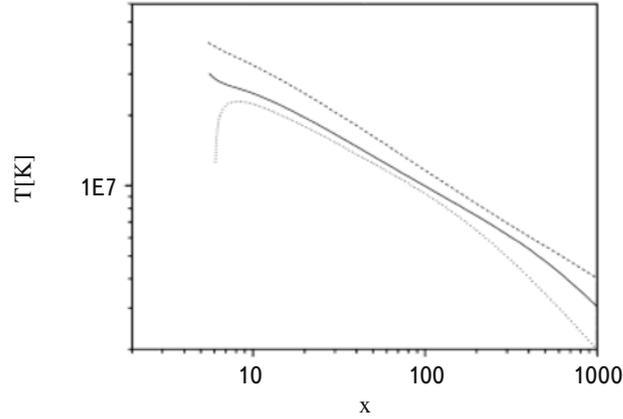

Fig. 5. The radial dependence of the temperature of the accretion disk for accretion rates $\dot{m} = 8$ (dotted curve), $\dot{m} = 30$ (smooth curve), and $\dot{m} = 150$ (dashed curve), and a viscosity parameter $\alpha = 0.1$.

## 6. Discussion and conclusions

We have obtained a unique solution for the structure of an advection accretion disk surrounding a nonrotating black hole for different values of the viscosity parameter and accretion rate. This solution is global, trans-sonic, and, for high $\dot{m}$ and $\alpha$, is characterized by a continuous transition of the disk from optically thick in the outer region to optically thin in the inner region.

The model, with a correct accounting for the transition between the optically thick and optically thin regions,

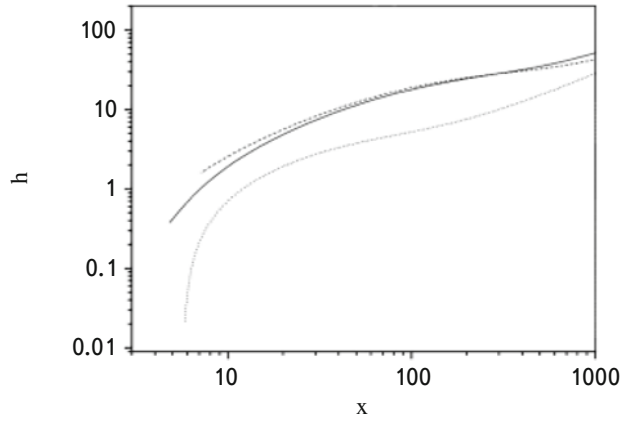

Fig. 6. The dimensionless half thickness of the accretion disk as a function of radius for the following parameters: $\dot{m}=10$ and $\alpha=0.01$ (dotted curve), $\dot{m}=50$ and $\alpha=0.01$ (smooth curve), and $\dot{m}=50$ and $\alpha=0.4$ (dashed curve).

reveals the existence of a temperature peak in the inner (optically thin) region. This peak might cause the appearance of a hard component in the spectrum, which could be observed. A high temperature in the inner region of an accretion disk may lead to the formation of electron-positron pairs and change the emission spectrum of the disk at energies of 500 keV and above.

We have shown that the existence and size of the optically thin region depend directly on the viscosity parameter $\alpha$. When $\alpha = 0.5$, a very substantial optically thin region is observed, when $\alpha = 0.1$ we have a slight optically thin region, and when $\alpha = 0.01$ no optically thin region is seen at all. This is because at very high $\dot{m}$ a large optical thickness is associated with a high density in the inner regions of the disk; it has a minimum at intermediate values of $\dot{m}$, and for $\alpha \leq 0.01$ this minimum turns out to be greater than unity.

We have also shown that the geometric thickness of the disk in this model depends substantially on the accretion rate and more weakly, on the viscosity parameter a.

We thank I. V. Igumenshchev for valuable advice.

This work was supported by the Russian Foundation for Basic Research (RFFI, grant No. 08-02-00491), the Program for support of leading scientific schools (grant NSH-2977.2008.2), and the Program of the Presidium of the Russian Academy of Sciences on the "Origin, structure, and evolution of objects in the universe."